\documentstyle[prl,aps]{revtex}
\draft
\begin{document}
\title{Kondo Screening in Gapless Magnetic Alloys}
\author{Valery I. Rupasov}
\address{Department of Physics, University of Toronto, Toronto,
Ontario, Canada M5S 1A7\\
and Landau Institute for Theoretical Physics, Moscow, Russia}
\date{\today}
\maketitle
\begin{abstract}
The low-energy physics of a spin-$\frac{1}{2}$ Kondo
impurity in a gapless host, where a density of band states
$\rho_0(\epsilon)=|\epsilon|^r/(|\epsilon|^r+\beta^r)$
vanishes at the Fermi level $\epsilon=0$, is studied
by the Bethe ansatz. The growth of the parameter
$\Gamma_r=\beta{\rm g}^{-1/r}$ (where ${\rm g}$ is an
exchange constant) is shown to drive the system ground
state from the Kondo regime with the screened impurity
spin to the Anderson regime, where the impurity spin is
unscreened, however, in a weak magnetic field $H$, it
exceeds its free value, $S_i(H)>\frac{1}{2}$, due to a
strong coupling to a band. It is shown also that a
sufficiently strong potential scattering at the impurity
site destroys the Anderson regime.
\end{abstract}

\pacs{PACS number: 72.15.Qm}

A growing body \cite{WF,BH,CF,CJ,I,GBI,BPH,IS} of
theoretical studies of unconventional magnetic alloys,
initiated by Withoff and Fradkin \cite{WF}, shows that
the standard picture of the Kondo effect in metals
\cite{H} should be fundamentally revised in the case of
so called ``gapless'' hosts, where an effective density
of band states vanishes precisely at the Fermi level
$\epsilon_F$ as $|\epsilon-\epsilon_F|^r$ with $r>0$.
``Poor-man's'' scaling arguments \cite{WF}, large-$N$ studies
\cite{WF,BH,CF}, and numerical renormalization group
calculations \cite{CJ,I,GBI,BPH} show that the Kondo
screening of the impurity spin in gapless systems
occurs only if an effective electron-impurity coupling
exceeds some critical value. Otherwise, an impurity
decouples from a band.

However, a Bethe ansatz (BA) analysis of the ground
state properties of an infinite-$U$ Anderson impurity
both in a BCS superconductor (a ``gapped'' Fermi
system) \cite{R1} and in a gapless host \cite{R2} have
shown no weak-coupling regime in the low-energy behavior
of the system. The ground state of an unconventional
Anderson system preserves basic characteristic features
of the metallic version. The appearance of a sufficiently
small gap or a pseudogap in a band dispersion results
only in some corrections to the standard solution
\cite{TW,PS}. In contrast to the metallic version, the
unconventional Anderson systems exhibit, however, a
nonuniversal behavior, that could explain discrepancies
between a BA solution and results of studies based on
scaling arguments.

In this Letter, we employ hidden integrability of
a spin-$\frac{1}{2}$ Kondo impurity in an unconventional
host \cite{R3} to explore the low-energy physics of gapless
systems, where an effective density of band states can
be modeled by
\begin{equation}
\rho_0(\epsilon)\equiv\frac{dk(\epsilon)}{d\epsilon}=
\frac{|\epsilon-\epsilon_F|^r}{|\epsilon-\epsilon_F|^r+\beta^r},
\;\;\;r>0.
\end{equation}
Here, $\epsilon_F$ is the Fermi energy, $k(\epsilon)$
is the inverse band dispersion, and the parameter $\beta$
characterizes the size of domain with a nonmetallic
behavior of $\rho_0(\epsilon)$.

In the Bethe ansatz approach to the theory of dilute
magnetic alloys \cite{TW,PS,AFL}, pioneered by Wiegmann
\cite{W1,W2} and Andrei \cite{A}, the spectrum of a free
host is alternatively described in terms of interacting
Bethe particles rather than in terms of free electrons
with spin ``up'' and ``down'', while an impurity plays
a role of an additional scattering center for Bethe
particles. Because of separation of the charge and spin
degrees of freedom, the spectrum of Bethe excitations,
in general, contains charge excitations, spin waves, and
their bound complexes.

In the Kondo model, an electron-impurity scattering is
energy independent, therefore the Bethe spectrum of the
model does not contain charge complexes. The ground state
of the system is composed of charge excitations and spin
waves only. The spin waves screen the impurity spin in
the zero-temperature limit $T\to 0$.

In the Anderson model, in contrast, the ground state
is composed of charge complexes, in which two charge
excitations are bound to a spin wave. Since charge
complexes are singlets, the spin of a spinless Anderson
impurity is naturally quenched in the ground state of
the system.

In unconventional hosts, scattering amplitudes acquire
an additional energy dependence because of an energy
dependent density of band states. In the Anderson
system, scattering amplitudes essentially depend on
energy already in a metallic host. An additional
dependence only renormalizes them slightly near the
Fermi level, that does not lead to any drastic changes
in the low-energy physics of the system in comparison
with the metallic version.

In the Kondo system, the situation is clear to be very
different. Since scattering amplitudes in the standard
Kondo model are energy independent, the appearance of
energy dependence in an unconventional host could really
lead to drastic changes in the physics of the system. As
in the Anderson models, the Bethe spectrum of the gapless
Kondo systems is shown to contain charge complexes. As
usual, only the simplest complexes contribute to the
ground state of the system. Therefore, one may restrict
a further consideration to (i) charge excitations, (ii)
spin waves, and (iii) the simplest charge complexes, in
which two charge excitations are bound to a spin wave.
To simplify our terminology, we use hereafter the terms
``particles'' and ``complexes'' to refer to charge
excitation and charge complexes, respectively.

One can propose two different physical scenarios of the
system behavior when the parameter $\beta$ increases
from its metallic value $\beta=0$. (i) One may expect
that at arbitrarily large $\beta$ the ground state of
a gapless system preserves basic characteristic features
of the standard Kondo model, so that the impurity spin
is screened. However, the Kondo temperature decreases
to extremely low temperatures at large $\beta$, and
the Kondo effect thus practically disappears. (ii) In
the second scenario, one may expect that the ground
states of a gapless system with a sufficiently large
$\beta$ and the metallic system are qualitatively
different, so that the impurity spin in a gapless host
is unscreened. In terms of BA language, it is obvious
that the only way to suppress the Kondo screening is
to reconstruct the ground state in such a way that all
spin waves are built into singlet complexes, as it takes
place in the Anderson systems.

To explore the low-energy physics of the system, we
derive the thermodynamic BA equations for renormalized
(fundamental) energies of Bethe excitations at a finite
temperature $T$ and then study the limit $T\to 0$. Solving
these equations at $T=0$, we find the ground state of the
system as a state, in which all states of Bethe excitations
with negative energies are filled out, while all states
with positive energies are empty.

To derive the thermodynamic BA equations one has to fix
the exponent $r$ in Eq. (1). In the cases of $r=\frac{1}{2}$,
$r=1$, and $r=2$, bare energies of complexes are {\em negative},
therefore they essentially affect the ground state properties
of the system. However, this is not sufficient yet to suppress
the Kondo screening. The second physical scenario presupposes
that complexes expel all particles and spin waves from the
ground state of the system. Only in this case, the Kondo
screening is suppressed completely.

The qualitative behaviors of the systems in all three
cases mentioned above are very similar, while the BA
mathematics in the $r=\frac{1}{2}$ and $r=1$ cases is
more tedious. To keep our mathematics as simple as possible,
we focus in this Letter on the $r=2$ case, which is, however,
of particular physical interest \cite{WF,BH,CF,CJ,I,GBI,BPH}.

We show that at a sufficiently large energy scale of
complexes $\Gamma_r=\beta{\rm g}^{-1/r}$, where ${\rm g}$
is an effective coupling constant, the renormalized energy
of particles is positive over the whole band. The growth of
the parameter $\Gamma_r$ drives thus the ground state of
the system from the Kondo type, in which spin waves screen
the impurity spin, to the Anderson type, in which all spin
waves are built into singlet complexes, and, therefore, the
impurity spin is unscreened, $S_i=\frac{1}{2}$.

Nevertheless, as in the case of the Anderson system,
we still deal with a {\em strong-coupling} regime of
our system. To clarify this point we study magnetic
properties of the ground state of the system in the
Anderson type regime. Despite the impurity spin is
unscreened and the magnetic susceptibility of the
impurity $\chi_i$ diverges in a weak external magnetic
field $H$, $\chi_i\sim H^{-1/3}$, the impurity does
not behave like a ``free'' localized magnetic moment.
Indeed, in a magnetic field, a part of complexes decay
into particles and spin waves. While spin waves disappear
to create a finite magnetization of a host, particles
bring a positive contribution to the unscreened impurity
spin. Thus, at $T=0$ the impurity spin in a magnetic field
exceeds its free magnitude, $S_i(H)>\frac{1}{2}$, that is
clear to have no analogy in a free impurity behavior.

An effective 1D Hamiltonian of the system is written in terms
of the Fermi operators $c_\sigma(\epsilon)$ which refer to
a band electron with spin $\sigma=\uparrow,\downarrow$ in an
$s$-wave state of energy $\epsilon$,
\begin{equation}
{\cal H}=\sum_{\sigma}\int\frac{d\epsilon}{2\pi}\,\epsilon
c^\dagger_\sigma(\epsilon)c_\sigma(\epsilon)
+\sum_{\sigma,\sigma'}
\int\frac{d\epsilon}{2\pi}
\frac{d\epsilon'}{2\pi}I(\epsilon,\epsilon')
c^\dagger_\sigma(\epsilon)
\left(\vec{\sigma}_{\sigma\sigma'}\cdot\vec{S}\right)
c_{\sigma'}(\epsilon')
\end{equation}
Here, $\vec{\sigma}$ are the Pauli matrices and $\vec{S}$
is the impurity spin operator. The electron energies and
momenta in Eq. (2) and hereafter are taken relative to
the Fermi values, which are set to be equal to zero. The
effective exchange coupling, $I(\epsilon,\epsilon')
=\frac{1}{2}I\sqrt{\rho_0(\epsilon)\rho_0(\epsilon')}$,
involve the exchange coupling constant $I$ and the
density of band states $\rho_0(\epsilon)$.

At an arbitrary density of band states, the model (1) is
diagonalized by the following BA equations \cite{R3}:
\begin{mathletters}
\begin{eqnarray}
\exp{(ik_jL)}\theta_{\frac{1}{2}}(u_j+1/{\rm g})
&=&\prod_{\alpha=1}^{M}\theta_1(u_j-\lambda_\alpha),\\
\theta_1(\lambda_\alpha+1/{\rm g})
\prod_{j=1}^{N}\theta_1(\lambda_\alpha-u_j)
&=&-\prod_{\beta=1}^{M}\theta_2(\lambda_\alpha-\lambda_\beta),
\end{eqnarray}
where $\theta_\nu(x)=(x-i\nu/2)/(x+i\nu/2)$, $k_j=k(\omega_j)$
and $\omega_j$ are electron momenta and energies, $N$ is
the total number of electrons on an interval of size $L$
and $M\leq N/2$ is the number of electrons with spin ``down''.
The eigenenergy $E$ and the $z$ component of the total
spin of the system $S^z$ are given by
\begin{equation}
E=\sum_{j=1}^{N}\omega_j,\;\;\;S^z=\frac{1}{2}+\frac{N}{2}-M,
\end{equation}
and an energy dependence of a charge ``rapidity''
$u_j=u(\omega_j)$ reads \cite{N1}
\begin{equation}
u(\omega)=\frac{2}{I}\frac{1}{\rho_0(\omega)}-\frac{3}{32}I\rho_0(\omega)
-\frac{1}{\rm g},
\end{equation}
where ${\rm g}^{-1}=2/I- 3I/32\simeq 2/I$ is an effective
coupling constant. The second term in $u(\omega)$ is much
smaller than the first one at all $\omega$, however, as it
will be clearly seen in what follows, this term plays a crucial
role in the low-energy physics of the system, and must be
kept.

In a metal, where $\beta=0$, and hence $\rho_0=1$ and
$u_j=0$, Eqs. (3) reduce to the BA equations of the
standard Kondo model. However, from the point of view of
the BA mathematics, they are similar to the BA equations
of the Anderson rather than the Kondo system. As in the
Anderson model, apart from particles with real energies
and momenta, the BA equations (3) admit also complexes
in which $2n$ charge excitations are bound to a spin complex
of order $n$. Thus, an energy dependence of charge rapidity
in an unconventional host essentially enriches the Bethe
spectrum of the Kondo system. As in the Anderson system,
only complexes of the lowest order, $n=1$, contribute to
the low-energy physics of the system. Therefore, we restrict
our consideration to the simplest complexes, in which
two charge excitations with complex energies
$\omega_\pm(\lambda)$,
\end{mathletters}
\begin{equation}
u(\omega_\pm)=\lambda\pm\frac{i}{2},
\end{equation}
and corresponding momenta $k_\pm(\lambda)\equiv
k[\omega_\pm(\lambda)]$ are bound to a spin wave
with a rapidity $\lambda$, provided that
$\mbox{Im}\, k_+(\lambda)>0$. A bare energy of a
complex
\begin{mathletters}
\begin{equation}
\xi_0(\lambda)=\omega_+(\lambda)+\omega_-(\lambda)=
-2\Gamma X(\lambda)-2\gamma x(\lambda),
\end{equation}
where $\Gamma=\beta/\sqrt{{\rm g}}$,
$\gamma=-\frac{3}{32}{\rm g}\Gamma=-\frac{3}{32}\beta\sqrt{{\rm g}}$,
and
\begin{eqnarray}
X(\lambda)&=&\sqrt{2}\frac{d}{d\lambda}
\left(\lambda+\sqrt{\lambda^2+1/4}\right)^{1/2},\\
x(\lambda)&\simeq&\frac{1}{(\lambda+1/{\rm g})\sqrt{\lambda}},
\end{eqnarray}
is negative at all $\lambda$. Here, we took into account
a smallness of the second term in Eq. (3d) which results
in the small second term in Eq. (5a), $\gamma\ll\Gamma$.
Moreover, for the latter we use only its asymptotic form
at $\lambda\gg 1$.

In the standard manner \cite{TW}, the thermodynamic BA
equations of our model for the renormalized energies of
particles, $\varepsilon(\omega)$, spin waves, $\kappa(\lambda)$,
and complexes, $\xi(\lambda)$, are found to be
\end{mathletters}
\begin{mathletters}
\begin{eqnarray}
\varepsilon(\omega)&=&\omega-\frac{1}{2}H
-\int_{-\infty}^{\infty}d\lambda\,
a_1[u(\omega)-\lambda]\,F[-\kappa(\lambda)]
+\int_{-\infty}^{\infty}d\lambda\,a_1[u(\omega)-\lambda]\,
F[-\xi(\lambda)],\\
\kappa(\lambda)&=&H+\int_{-\infty}^{\infty}d\omega\,u'(\omega)\,
a_1[\lambda-u(\omega)]\,F[-\varepsilon(\omega)]
+\int_{-\infty}^{\infty}d\lambda'\,
a_2(\lambda-\lambda')\,F[-\kappa(\lambda')],\\
\xi(\lambda)&=&\xi_0(\lambda)
+\int_{-\infty}^{\infty}d\omega\,u'(\omega)\,a_1[\lambda-h(\omega)]
F[-\varepsilon(\omega)]
+\int_{-\infty}^{\infty}d\lambda'\,a_2(\lambda-\lambda')
F[-\xi(\lambda')].
\end{eqnarray}
Here, $F[f(x)]\equiv T\ln{\{1+\exp{[f(x)/T]}\}}$,
$u'(\omega)=du/d\omega$, $a_\nu(x)=(2\nu/\pi)(\nu^2+4x^2)^{-1}$,
and $H$ is an external magnetic field.

In the zero-temperature limit, $T\to 0$, all states
with negative energies must be filled out, while all
states with positive energies must be empty. In the
absence of a magnetic field, the magnetization of a
host must be equal to zero. This implies that at $H=0$
the number of particles in the system is twice bigger
than the number of spin waves. Therefore, at $H=0$ the
ground state is composed of charge complexes only if
\end{mathletters}
\begin{equation}
\kappa(\lambda)>0,\; \lambda\in(-\infty,\infty);\;\;\;
\varepsilon(\omega)>0,\; \omega\in(-\epsilon_F,\epsilon_F),
\end{equation}

The energy of spin waves is easily seen from Eq. (6b) to
be positive, provided $\varepsilon(\omega)>0$. Therefore,
in the limit $T\to 0$, the conditions (7) reduce to
$\varepsilon(\omega)>0$ at $H=0$, where
\begin{mathletters}
\begin{equation}
\varepsilon(\omega)=\omega-\frac{1}{2}H
-\int_{-\infty}^{\infty}d\lambda\,a_1[u(\omega)-\lambda]\,
\xi(\lambda),
\end{equation}
and the energy $\xi(\lambda)$ is found from the equation
\begin{equation}
\xi(\lambda)=\xi_0(\lambda)-
\int_{-\infty}^{\infty}d\lambda'\,a_2(\lambda-\lambda')\,\xi(\lambda').
\end{equation}
Inserting a solution of Eq. (8b) into Eq. (8a), one
obtains
\end{mathletters}
\begin{mathletters}
\begin{equation}
\varepsilon(\omega)=\omega-\frac{1}{2}H
+2\Gamma\int_{-\infty}^{\infty}d\lambda\,s[u(\omega)-\lambda]\,
X(\lambda)
+2\gamma\int_{-\infty}^{\infty}d\lambda\,s[u(\omega)-\lambda]\,
x(\lambda),
\end{equation}
where $s(x)=[2\cosh{(\pi x)}]^{-1}$.

As $\omega\to 0$ the rapidity $u(\omega)\to\infty$, and we
find
\begin{equation}
\varepsilon(\omega)\simeq\omega-\frac{1}{2}H+|\omega|+
\gamma\frac{|\omega|^3}{\Gamma^3}
+{\cal O}\left(|\omega|^5\right).
\end{equation}
At $\omega<0$, the first and third terms are cancelled,
while the forth term determines a small positive
contribution to the particle energy. In a sufficiently
weak magnetic field, the function $\varepsilon(\omega)$
is negative in a small domain between points $\Omega_{-}
\simeq-\Gamma(H/2\gamma)^{1/3}$ and $\Omega_+\simeq H/4$,
where $\Omega_{\pm}$ are found from the equation
$\varepsilon(\Omega_{\pm})=0$.

As $\omega\to\infty$ the function $\varepsilon(\omega)
\simeq\omega+\mbox{const}$, and hence it has the third zero,
at some point $\omega=\Omega<0$. The critical magnitudes of
the parameters at which particles and spin waves disappear
from the ground state of the system are clear to be determined
by the condition $\Omega=-\epsilon_F$, or
\end{mathletters}
\begin{equation}
2{\cal G}\int_{-\infty}^{\infty}d\lambda\,
s({\cal G}^2 -\lambda)\left[X(\lambda)+
\frac{3}{32}{\rm g}\,x(\lambda)\right]=1,
\end{equation}
where ${\cal G}=\Gamma_{\rm cr}/\epsilon_F$. In the absence
of the second term in the brackets, the left-hand side
of Eq. (10) is less than $1$, but it is asymptotically
very close to $1$ already at ${\cal G}\geq 1$. Therefore,
an estimate $\Gamma_{\rm cr}\simeq\epsilon_F$ works very well
at all reasonable values of the coupling constant ${\rm g}$.

Thus, at $\Gamma<\Gamma_{\rm cr}$, the ground state of the
system is composed of particles, spin waves, and
complexes. The complexes essentially affect the ground
state properties, however, as in a metal, ``free'' spin
waves, unbuilt into complexes, screen completely the
impurity spin in the absence of a magnetic field, $S_i=0$.
At $\Gamma>\Gamma_{\rm cr}$, the scattering on complexes
renormalizes the energy of particles to positive values
over the whole band. In other words, particles and spin
waves are completely expelled from the ground state of
the system, which is composed now of singlet complexes
only. Therefore, the impurity spin is unscreened and
equal to its free magnitude, $S_i=\frac{1}{2}$, in the
absence of a magnetic field.

Despite the impurity spin is unscreened, the impurity
is easily seen to be strongly coupled to a band. In the
continuous limit \cite{TW,PS,AFL}, Eqs. (3) for the
system ground state take the form of integral equation
for the densities of states of particles, $\rho(\omega)$,
and complexes, $\sigma(\lambda)$,
\begin{mathletters}
\begin{eqnarray}
\rho(\omega)&=&\frac{1}{L}\,\delta[u(\omega)+1/{\rm g}]
+\frac{1}{2\pi}\rho_0(\omega)
-u'(\omega)\int_{-\infty}^{\infty}d\lambda\,
a_1[u(\omega)-\lambda]\sigma(\lambda)\\
\sigma(\lambda)&=&\frac{1}{L}\Delta(\lambda+1/{\rm g})+
\frac{1}{2\pi}\frac{dp(\lambda)}{d\lambda}
-\int_{-\infty}^{\infty}d\lambda'
a_2(\lambda-\lambda')\sigma(\lambda')
-\int_{\Omega_-}^{\Omega_+}d\omega\,
a_1[\lambda-u(\omega)]\rho(\omega).
\end{eqnarray}
Here, $p(\lambda)=k_+(\lambda)+k_-(\lambda)$ is
the momentum of a complex, while the functions
$\delta[u(\omega)]=u'(\omega)a_\frac{1}{2}[u(\omega)]$
and $\Delta(\lambda)=a_{\frac{3}{2}}(\lambda)-a_1(\lambda)-
a_{\frac{1}{2}}(\lambda)$ describe the scattering of particles
and complexes at the impurity site. Separating the densities
into the host and impurity parts, $\rho(\omega)=
\rho_h(\omega)+L^{-1}\rho_i(\omega)$, we find for the
impurity spin
\end{mathletters}
\begin{equation}
S_i=\frac{1}{2}+
\frac{1}{2}\int_{\Omega_-}^{\Omega_+}d\omega\rho_i(\omega).
\end{equation}
Since complexes carry no spin, they do not contribute to
the impurity spin. At $H=0$, $\Omega_{\pm}=0$, and the
impurity spin is equal to its free magnitude,
$S_i=\frac{1}{2}$. In the presence of an arbitrarily weak
field, $\Omega_{\pm}\neq 0$, the unscreened impurity spin
acquires a positive contribution due to a strong coupling
to the host band. In a weak field, the last term in the
right-hand side of Eq. (11b) is small and can be omitted
in the zero-order computations. Then, taking into account
that $\Delta(\lambda)\sim\lambda^{-4}$ as $\lambda\to\infty$,
one obtains $\rho_i(\omega)\simeq\delta[u(\omega)]$ as $H\to 0$,
and the impurity spin is found to be
\begin{equation}
S_i=\frac{1}{2}+\frac{1}{8\pi}\left(\frac{H}{2\gamma}\right)^{2/3}
+{\cal O}(H^2).
\end{equation}
While the magnetic susceptibility of the host vanishes,
$\chi_h\sim H^2$, the impurity susceptibility,
$\chi_i\sim H^{-1/3}$, diverges as $H\to 0$. Thus, the
Nozi\`eres theory \cite{N} based on the Fermi liquid approach
is not generalized to the case of a gapless host.

Finally, it should be noted that the unscreened impurity
spin regime is very sensitive to a potential (spin independent)
scattering at the impurity site. While in a metal potential
scattering does not play any essential role \cite{AFL}, in
a gapless host, it is essentially affect the expression for
charge rapidity $u(\omega)$. If a potential scattering with
a coupling constant $V$ is taken into account \cite{R3},
the expression (3d) takes the form
$$
u(\omega)=\frac{2}{I}[\rho^{-1}_0(\omega)-1]
+\frac{\left(V+\frac{1}{4}I\right)
\left(V-\frac{3}{4}I\right)}{2I}[\rho_0(\omega)-1]
$$
At $V=0$ the second positive term has been shown to result
in expelling particles from the system ground state. It is
clear that if $V$ lies outside the interval $(-I/4,3I/4)$,
the sign of this term is negative, that immediately destroys
the Anderson (unscreened impurity spin) regime.

I am thankful to S. John for stimulating discussions.


\begin{references}

\bibitem[*]{ea}
Electronic address: rupasov@physics.utoronto.ca

\bibitem{WF}
D. Withoff and E. Fradkin, \prl {\bf 64}, 1835 (1990).

\bibitem{BH}
L. S. Borkowski and P. J. Hirschfeld, \prb {\bf 46},
9274 (1992).

\bibitem{CF}
C. R. Cassanello and E. Fradkin, \prb {\bf 53}, 15 079 (1996);
{\em ibid.} {\bf 56}, 11 246 (1997).

\bibitem{CJ}
K. Chen and C. Jayaprakash, J. Phys.: Condens. Matter,
{\bf 7}, L491 (1995).

\bibitem{I}
K. Ingersent, \prb {\bf 54}, 11 936 (1996).

\bibitem{GBI}
C. Gonzalez-Buxton and K. Ingersent, \prb {\bf 54},
15 614 (1996); {\em ibid.} {\bf 57}, 14 254 (1998).

\bibitem{BPH}
R. Bulla, Th. Pruschke, and A. C. Hewson, J. Phys.:
Condens. Matter {\bf 9}, 10463 (1997).

\bibitem{IS}
K. Ingersent and Q. Si, preprint cond-mat/9810226.

\bibitem{H}
A. C. Hewson, {\em The Kondo Problem to Heavy Fermions},
(Cambridge University Press, Cambridge, 1993)

\bibitem{R1}
V. I. Rupasov, \prl {\bf 80}, 3368 (1998).

\bibitem{R2}
V. I. Rupasov, \prl {\bf 82}, 839 (1999).

\bibitem{TW}
A. M. Tsvelick and P. B. Wiegmann, Adv. Phys. {\bf 32}, 453 (1983).

\bibitem{PS}
P. Schlottmann, Phys. Rep. {\bf 181}, 1 (1989).

\bibitem{AFL}
N. Andrei, K. Furuya, and J. H. Lowenstein, \rmp {\bf 55}, 331 (1983).

\bibitem{R3}
V. I. Rupasov, preprint cond-mat/9812086;
to appear in Phys. Lett. A (1999).

\bibitem{W1}
P. B. Wiegmann, JETP Lett. {\bf 31}, 364 (1980)
[Pis'ma Zh. Eksp. Teor. Fiz. {\bf 31}, 397 (1980)]

\bibitem{W2}
P. B. Wiegmann, Phys. Lett. A {\bf 80}, 163 (1980).

\bibitem{A}
N. Andrei, \prl {\bf 45}, 379 (1980).

\bibitem{N1}
We omitted here a weak dependence of $u(\omega)$ on
the real part of the self-energy of electron-impurity
scattering $\Sigma'(\omega)=P\int(d\epsilon/2\pi)
\rho_0(\epsilon)/(\omega-\epsilon)$ \cite{R3}.

\bibitem{N}
P. Nozi\`eres, J. Low-Temp. Phys. {\bf 17}, 13 (1974).
\end{references}
\end{document}